          [2001/04/25 1.1 (PWD)]
\documentclass[a4paper,twocolumn]{esapub} % European paper
\usepackage{natbib}
\usepackage{graphicx}

\title{Preliminary \textit{INTEGRAL} Analysis of GRB\,040106}
\author{L. Moran}
\author{L. Hanlon}
\affil{Department of Experimental Physics, University
   College Dublin, Ireland, lynn.moran@ucd.ie, lorraine.hanlon@ucd.ie}
\author{A. von Kienlin}
\affil{Max-Planck-Institut f\"{u}r extraterrestrische Physik, 85748
  Garching, Germany, azk@mpe.mpg.de}
\author{B. McBreen}
\author{S. McBreen}
\author{S. McGlynn}
\author{J. French}
\affil{Department of Experimental Physics, University
   College Dublin, Ireland, brian.mcbreen@ucd.ie, smcbreen@bermuda.ucd.ie,
   jfrench@bermuda.ucd.ie, smcglynn@bermuda.ucd.ie}
\author{ R. Preece}
\author{Y. Kaneko}
\affil{Department of Physics, University of Alabama at
 Huntsville, USA, Robert.Preece@msfc.nasa.gov, Yuki.Kaneko@msfc.nasa.gov}
\author{O.R. Williams}
\author{K. Bennett}
\affil{Science Operations and Data Systems Division of
   ESA/ESTEC,SCI-SDG, NL-2200 AG Noordwijk, The Netherlands,
   owilliam@rssd.esa.int, kbennett@rssd.esa.int}
\author{R. Marc Kippen}
\affil{Space and Remote Sensing Sciences, Los Alamos National
 Laboratory, USA, rmkippen@lanl.gov}

\begin{document}


\maketitle
\begin{abstract}
On January 6$^{th}$ 2004, the IBAS burst alert system
    triggered the 8$^{th}$ gamma--ray burst (GRB) to be detected by the
    \textit{INTEGRAL} satellite. The position was determined
    and publicly distributed within 12\,s, enabling ESA's \textit{XMM--Newton} to take
    advantage of a ToO observation just 5\,hours later during which the
    X--ray afterglow was detected. Observations at
    optical wavelengths also revealed the existence of a fading
    optical source. The GRB is $\sim$\,52\,s long with 2 distinct
    peaks separated by $\sim$\,24\,s. At gamma--ray
    energies the burst was the weakest detected by \textit{INTEGRAL} up to
    that time with a flux in the 20\,-\,200\,keV band of 0.57\,photons\,cm$^{-2}$\,s$^{-1}$. Nevertheless, it was possible
to determine its position and extract spectra and fluxes. Here we
    present light curves and the results of imaging, spectral and temporal
    analyses of the prompt emission and the onset of the afterglow from \textit{INTEGRAL} data.

\end{abstract}

\section{Introduction}
Gamma--ray bursts are an amazingly energetic phenomenon, capable of an
isotropic output of order 10$^{52}$-10$^{54}$\,erg in a few
seconds. Although first detected in the late 1960s, significant
progress has mostly been achieved in the last dozen years. That GRBs are extra--galactic in origin
was suggested by the isotropic distribution of GRBs observed by BATSE on board the Compton Gamma--Ray Observatory \citep{mee1992,fish1994}. The discovery
by BeppoSAX of afterglows in the X--ray \citep{costa:1997} and subsequent discoveries at
optical \citep{vanp:1997} and radio \citep{frail:1997} wavelengths have led to redshift
measurements \citep{metz:1997} for $\sim$\,40 bursts ranging from
$z = 0.168-4.5$. A theory of gamma--ray bursts must provide a
mechanism capable of releasing enormous quantities of non--thermal
energy by compact sources at cosmological distances.

Although not built as a GRB oriented mission, \textit{INTEGRAL} has a burst alert
system (IBAS) and the two main instruments on board have coded masks, a wide
field of view (FoV), cover a wide energy range (15\,keV\,-\,8\,MeV) and offer high resolution capabilities in imaging
(IBIS) and spectroscopy (SPI). IBAS carries out rapid localisations
for GRBs incident on the IBIS detector with precision of a few
arcminutes \citep{vonk2003}. The public distribution of these co--ordinates enables
multi--wavelength searches for afterglows at lower energies. \textit{INTEGRAL}
data of the prompt emission in combination with the early
multi--wavelength studies offers the best currently available probe of
the origin of these transient phenomena.

In \S 2 observations and imaging analysis of GRB\,040106 are presented. \S 3 describes the spectral analysis
methods utilised and the results obtained from analysis of SPI data. A
brief account of the temporal analysis is presented in \S 4.

\begin{figure}
\centering
\includegraphics[width=0.9\linewidth]{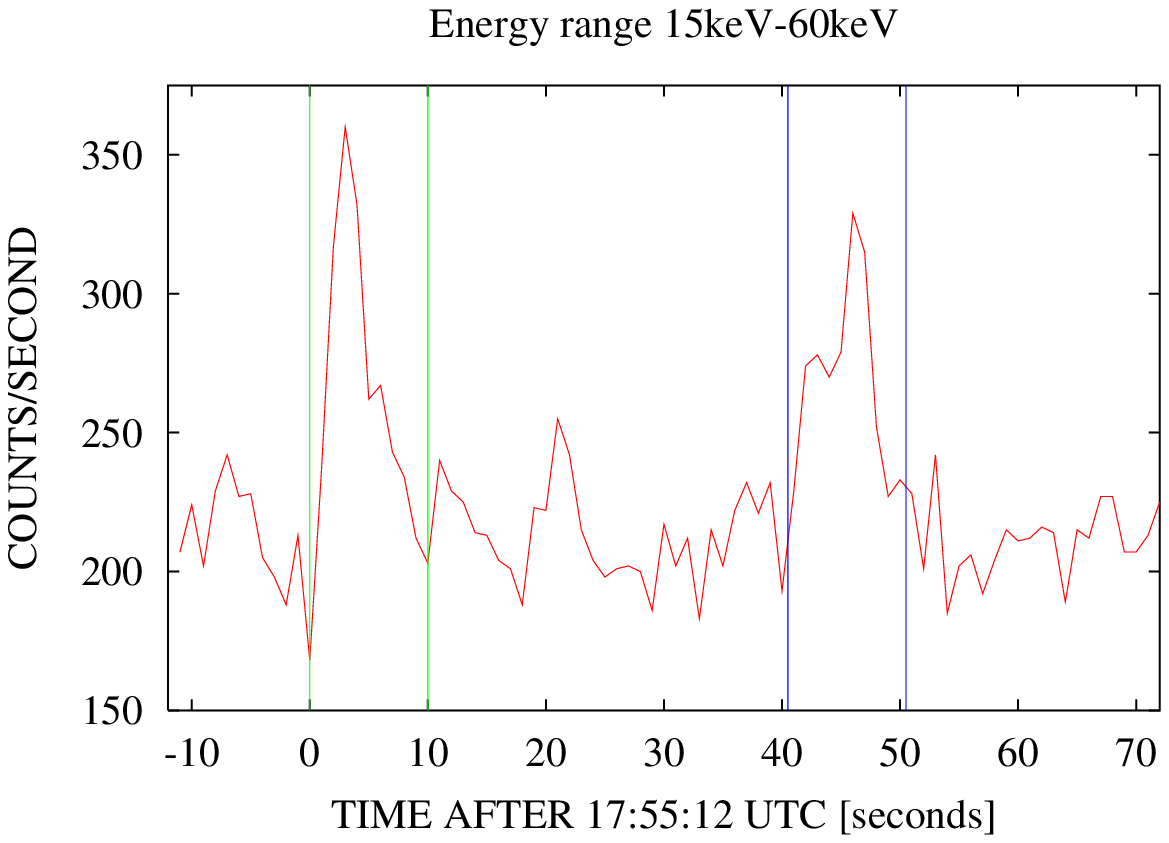}
\includegraphics[width=0.9\linewidth]{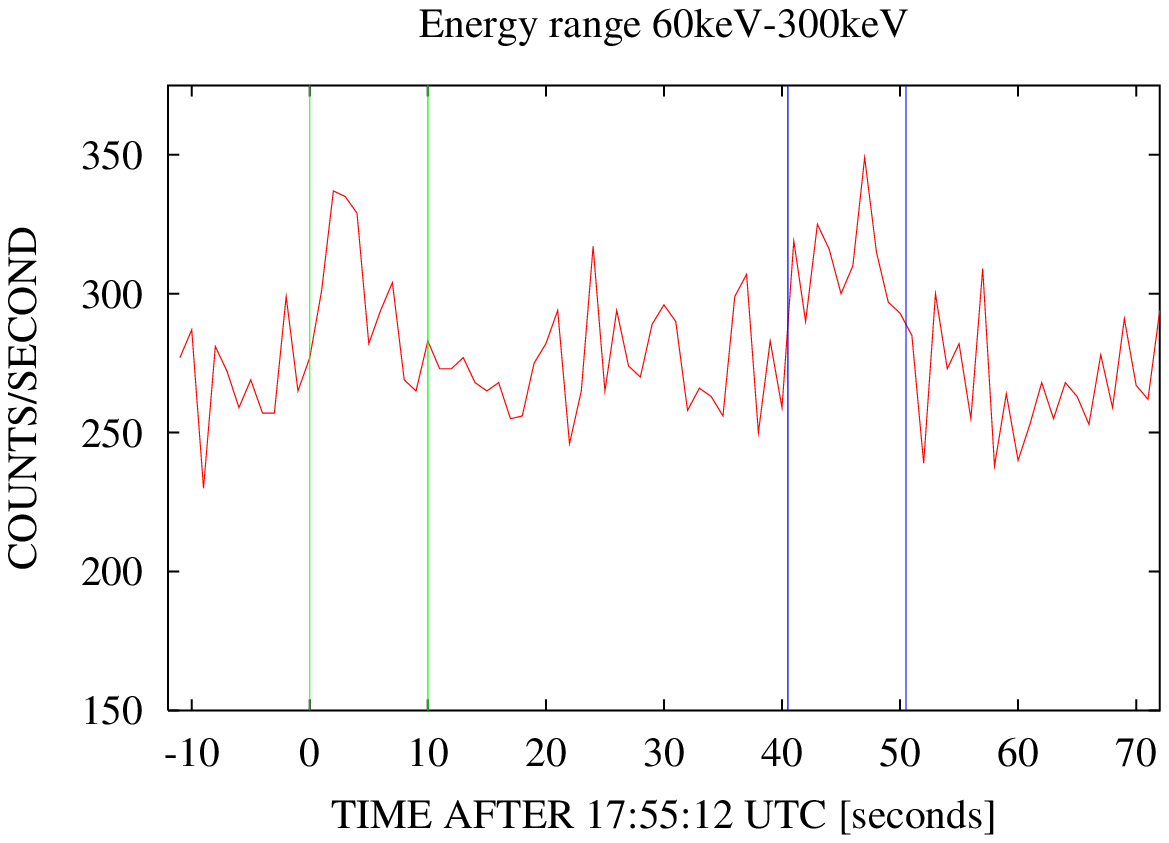}
\caption{ISGRI light curves of GRB\,040106 in 2 energy ranges (upper panel) 15\,-\,40\,keV and (lower panel) 40\,-\,200\,keV}\label{ibis_lc}
\end{figure}

\section{Detection and Localisation}

GRB\,040106 was detected at 17:55:11\,UTC on January 6$^{th}$ 2004. Lying at an
off--axis angle of 10.5$^\circ$, the GRB was visible in the partially
coded FoV of IBIS and of SPI, but was outside the
FoV of the two monitoring instruments, JEM--X and the OMC.  This GRB
falls into the class of long GRBs, lasting approximately 52\,s with
two prominent peaks separated by a quiescent period of $\sim$\,24\,s. The ISGRI
detector of IBIS, which is most sensitive between 15\,-\,300\,keV, is
used to produce the light curves of GRB\,040106 shown in
Fig. \ref{ibis_lc}. The imaging capabilities of SPI are due to a coded mask comprising of 127 tungsten elements, with a
thickness of 30\,mm, placed at a distance of $\sim$\,1.7\,m from the
detection plane, providing an angular resolution of 2.5$^\circ$ \citep{ved2003}. Due
to the weak nature of this burst, data from the two time intervals around the prominent
peaks of emission (Fig. \ref{ibis_lc}) were combined to enable SPI to determine a position
for the GRB in the range 20\,-60\,keV. The position for GRB\,040106
extracted from the SPI data is $\alpha_{J2000}$\,=\,11$^h$ 52$^m$ 51.12$^s$,
$\delta_{J2000}$\,=\,-46$^\circ$ 47$'$ 13.2$''$, which is 5.7$'$ off the
position determined from the ISGRI data.

The IBAS alert \citep{mere2004} was automatically distributed approximately 12\,s after
the burst start time with a positional uncertainty of only
3$'$. An \textit{XMM--Newton} ToO observations began a 45000\,s exposure at
23:05\,UT. A bright source was visible in the 30\,ks
Quick--Look--Analysis \citep{ehle2004}, 0.9$'$ from the IBAS
position \citep{tedds2004}. At optical wavelengths the REM Telescope
at the European Southern Observatory reported no new sources
\citep{palaz2004} detected during an observation at 04:45 on January 7$^{th}$. The Swope Telescope at Las Campanas Observatory identified
two faint sources as candidate optical afterglows of GRB\,040106
\citep{price2004}, but these were later observed to have retained the
same flux to $\pm$\,0.1\,magnitude and hence are not candidate optical
counterparts to GRB\,040106 \citep{fox2004}. Observations with ESO's
New Technology Telescope on two consecutive nights immediately after
the GRB found one source demonstrating fading behaviour from 22.4\,mag
to 23.7\,mag \citep{mase2004}, consistent the IBIS error circle and
hence identifying a likely optical afterglow. Radio observations on
January 10$^{th}$ with the Australia Telescope Compact Array (ATCA) at
a frequency of 8.6\,GHz detected a 160\,$\mu$Jy source with a
5\,$\sigma$ significance 1.8$'$ from the centre of the
\textit{XMM--Newton} error circle \citep{wier2004}. However this source is not
consistent with any of the optical sources reported and was not
observed on January 21$^{st}$ with the Very Large Array
\citep{frail2004}.

\section{Spectral Analysis with SPI}

We have investigated the spectral evolution of GRB\,040106 with
 SPI. The two bright peaks indicated in Fig. \ref{ibis_lc}, were selected for analysis. The first interval is 7\,s long starting at the
very beginning of the burst 17:55:11\,UTC, and the second begins 34\,s
later and lasts for 12\,s. Spectra for each of these intervals were
 extracted at the XMM--Newton afterglow position. 

Several background handling approaches were considered. The `CONSTANT'
  background applies the same background to each detector. The default
  value of 10$^{-7}$ counts/det/sec/keV was chosen for this
  analysis. The `ACS' method assumes that the background
  follows the time variation of the total number of Anti--Coincidence
  Shield counts, scaled appropriately per detector in the spectral extraction. There is good agreement between the results obtained
  with ACS and CONSTANT backgrounds, when the GRB is in SPI's FoV it
  is not detectable by the ACS. The `DFEE' background utilises the count rates of the individual
  detectors to create a time variable model background. The `SPIOFFBACK'
 program calculates the relative intensity of the background in each
 detector by examining the counts for a period in the same pointing,
  but excluding the time around GRB and scaling to the GRB duration
  thus all sources in the FoV, except the GRB, are subtracted as
  background. The DFEE results yield harder
  indices for both peaks than the other three methods and produces a
  hard excess above a few hundred keV. Therefore, unsurprisingly,
  since it assumes all sources are constant, this is not a suitable
  background choice for GRB analysis. The SPIOFFBACK results are
  consistent with the other methods and because its purpose is to
  remove all other sources in the FoV, it is the most suitable method
  for GRB studies. In this case a 40\,minute period
  in the same science window as the GRB but before the trigger, is
  used to generate the background.

\begin{figure}
\centering
\includegraphics[angle=270,width=0.9\linewidth]{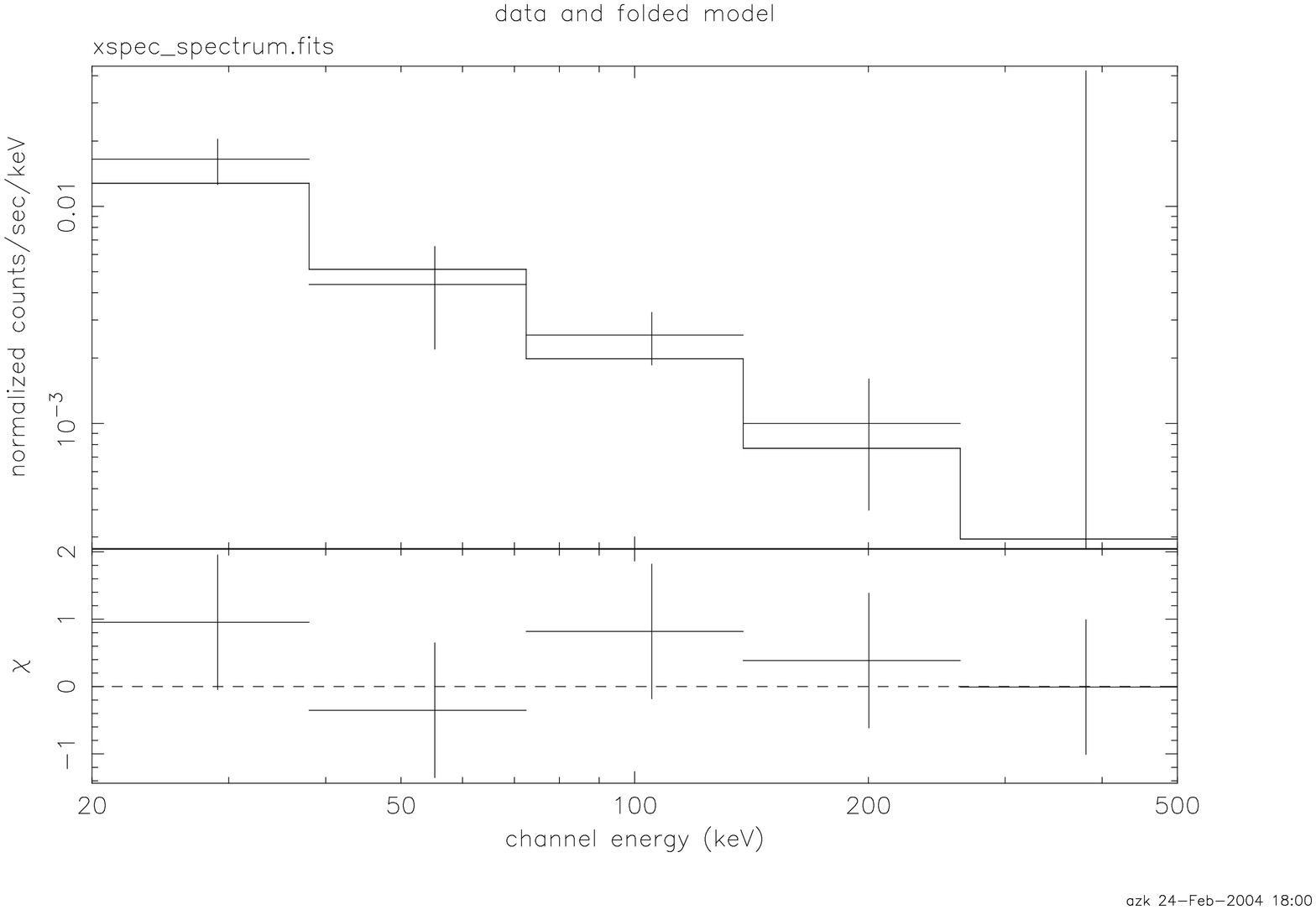}
\includegraphics[angle=270,width=0.9\linewidth]{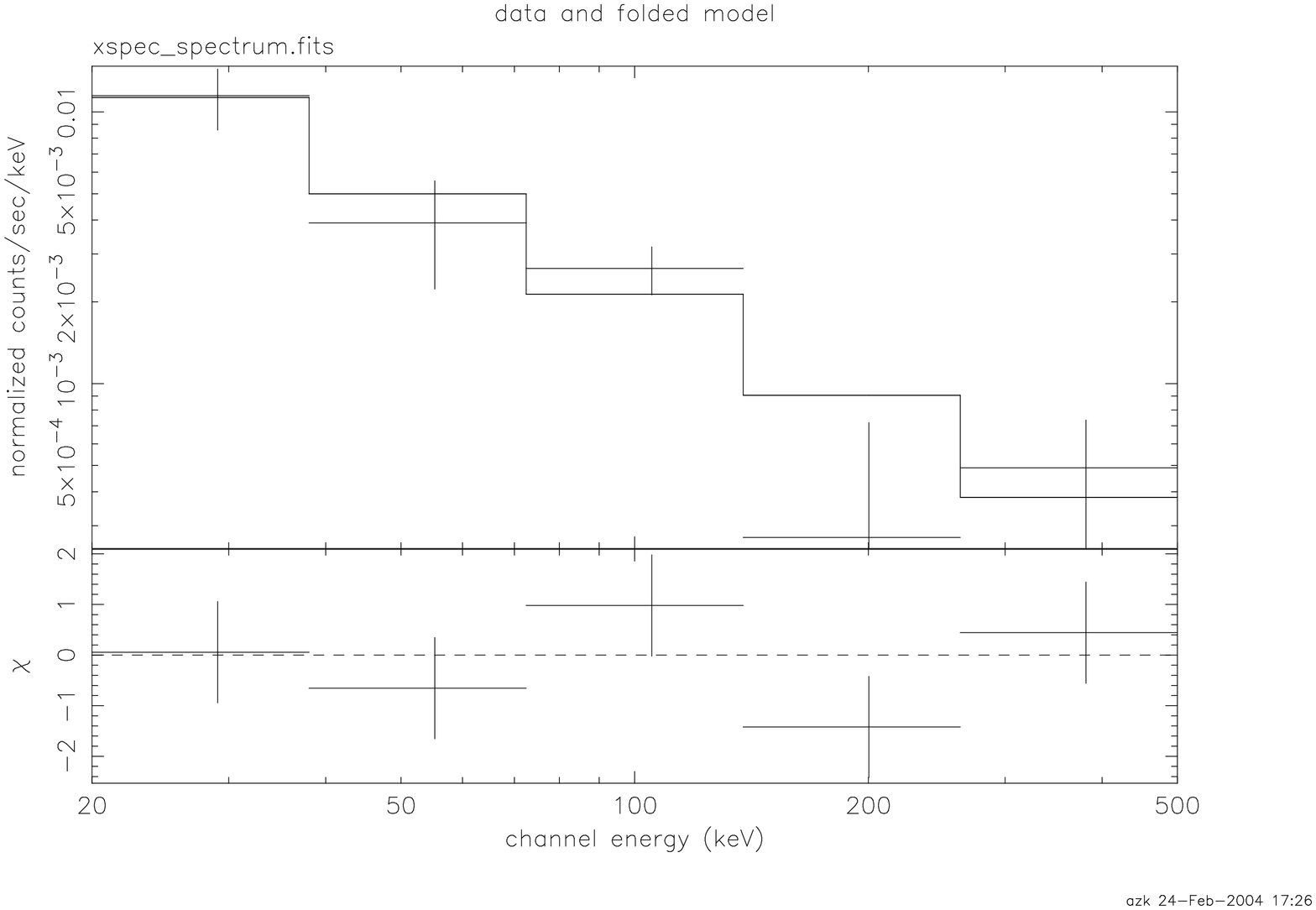}
\caption{Spectra of the first and second peaks of GRB\,040106 fit with
a single power law model}\label{spec}

\end{figure}

The single events detected by SPI, corrected for intrinsic deadtimes and
 telemetry gaps, are binned into 5 equally spaced logarithmic energy
 bins in the 20\,keV to 200\,keV range for each of the chosen
 intervals. The SPIROS  (SPI Iterative Removal of Sources, \cite{skin2003}) software
 package is used for spectral extraction, while XSPEC\,11.2 is used
 for model fitting. The spectrum for each of the two intervals is best
 fit by a single power law model (Fig. \ref{spec}). The photon
 indices, normalisation values and fluxes obtained are presented in
 Table \ref{tab_spec}. Inspection of the IBIS/ISGRI light curves in low and high energy
 bands (Fig. \ref{ibis_lc}) suggests that the second peak is harder
 than this first. Errors quoted are 1 parameter of interest at 67\%
 confidence level. Spectral analysis yields a photon index of
 $1.47^{+0.59}_{-0.52}$ for the first peak, but
 $1.32^{+0.34}_{-0.31}$ for the second peak confirming that the second
 peak is harder, while the flux remains approximately constant
 $\sim6.1 \times10^{-8} erg cm^{-2} s^{-1}$ for both peaks. The same analysis was conducted for multiple events incident on the
SPI detectors, but yielded no significant improvement to the fit.

\begin{table}
\caption{Spectral analysis of GRB\,040106 with SPI for two
  intervals around the prominent peaks of emission}
\begin{center}
\begin{tabular}{l|lc}
\hline\\
\textit{Interval} & \textit{Parameter} & SPI\\
\hline\hline\\
1$^{st}$ peak & Photon Index & 1.47$^{+0.59}_{-0.52}$\\
& Normalisation & 1.67 \\
& $^*$Flux (erg\,cm$^{-2}$\,s$^{-1}$) & 5.9$\times$10$^{-8}$ \\
\hline\\
2$^{nd}$ peak & Photon Index & 1.32$^{+0.34}_{-0.31}$ \\
& Normalisation & 0.90 \\
& $^*$Flux (erg\,cm$^{-2}$\,s$^{-1}$) &  6.1$\times$10$^{-8}$ \\
\hline
\hline

\end{tabular}
\end{center}
$^*$flux is measured in the energy range 20\,-\,200\,keV
\label{tab_spec} 
\end{table}

\begin{center}
\begin{table}
\caption{Timing analysis of GRB\,040106}
\begin{center}
\begin{tabular}{l|ccc}
\hline\\
\textit{Parameters} & \textit{t$_r$ (s)} & \textit{t$_f$ (s)} & \textit{FWHM}\\
\hline\hline\\
1$^{st}$ Pulse & 2.4 & 2.8 & 4.5\\
2$^{nd}$ Pulse & 4.1 & 5.8 & 10.1\\

\hline
\hline

\end{tabular}
\label{tab_tim}
\end{center} 
\end{table}
\end{center}

\section{Temporal Analysis}

Table \ref{tab_tim} summarises the results of denoising GRB\,040106
using a wavelet analysis and the temporal properties of the two
significant pulses. The duration (T90) of the burst is 52\,s and there
is a time interval, $\Delta$T, of 42\,s between the peaks. The pulse properties of pulses in short and long GRBs are
consistent with lognormal distributions
\citep{Quil2002}. Furthermore the pulse properties and time
intervals between pulses are related to T90 \citep{mcbreen2002a} and
presented in the form of a set of timing diagrams. The time interval
and properties of the two pulses, $\Delta$T, the rise time (t$_r$),
fall time (t$_f$)
and FWHM of this GRB are are proportional to T90 and fit well on the
timing diagrams. A long, weak GRB generally has slow pulses well separated from each other,
rather than fast pulses close together. There is no satisfactory explanation
of this phenomenon but it is probably related to a low value of the
bulk Lorentz factor $\Gamma$ and a viscous accretion disk surrounding a black hole \citep{fry1999}.

\section{Discussion \& Conclusions}
On December 19$^{th}$ 2003, SPI's detector \#2 was confirmed dead
after several attempts to revive it met with no success. As yet a new
redistribution matrix has not been released to take account of this
change in SPI's response. GRB\,040106 was incident on detector \#2 and
the surrounding detectors, so an improvement in the fit is expected when
the new matrix is issued.

The results of spectral analysis of GRB\,040106 confirm that the
second peak is harder than the first and that it is well fit by a
single power--law model of photon index $\alpha \sim$ -1.3. It is
unlikely that this is an X--ray rich GRB with a peak energy at or
below the low end of the SPI detector sensitivity
(i.e. $\sim$\,20\,keV) since the spectral index would 
correspond to an unusually hard value for the high--energy index
above the spectral turnover \citep{pree2000}. It is more likely that
the weakness of this GRB washes out evidence for a spectral break
at more typical energies of a few hundred keV.
There is no evidence in the IBIS/ISGRI light curve for soft
extended or delayed emission such as that observed by, for example, SIGMA/GRANAT
in GRB\,920723 \citep{buren1999} or by HETE--II in GRB\,021211
\citep{crew2003}. The temporal decay of the 2$^{\rm nd}$ peak is
consistent with a power--law of slope $\beta = -0.3\pm 0.7$ which
may indicate the presence of a high--energy afterglow, due to
external shocks, during the burst itself. However, the data are
not sufficiently constrained to indicate fast or slow, or
radiative or adiabatic, cooling in the synchrotron shock model
\citep{gib2002,piro2004}. Further analysis and comparison with XMM--Newton
results are on--going \citep{moran2004}.

\section*{Acknowledgments}

We would like to thank the staff of the INTEGRAL Science Data Centre
for their support with this work.

% The following bibliography was produced with
%   \bibliographystyle{aa}
%   \bibliography{esapub}
% The results are inserted directly here to simplify
% the demonstration.

\bibliography{/home/lmoran/WORK/WRITING/refs}
\bibliographystyle{aa}
\end{document}